\documentclass[conference]{IEEEtran}
\IEEEoverridecommandlockouts
\usepackage{cite}
\usepackage{amsmath,amssymb,amsfonts}
\usepackage{mdwmath}
\usepackage{algorithmic}
\usepackage{graphicx}
\usepackage{textcomp}
\usepackage{bm}
\usepackage{algorithm}
\usepackage[font=footnotesize]{caption}
\usepackage{subcaption}
\usepackage{algorithmic}
\usepackage{multirow}
\usepackage{hyperref}
\usepackage{comment}
\usepackage{xcolor}
\def\BibTeX{{\rm B\kern-.05em{\sc i\kern-.025em b}\kern-.08em
    T\kern-.1667em\lower.7ex\hbox{E}\kern-.125emX}}
\begin{document}

\title{Reducing AoI and Improving Throughput for NOMA-assisted SGF Systems: A Hierarchical Learning Approach\\
}

\author{Yuqin~Liu,~\IEEEmembership{Student Member,~IEEE,}  Mona~Jaber,~\IEEEmembership{Senior~Member,~IEEE,} Yan~Liu,~\IEEEmembership{Member,~IEEE,} \\and Arumugam~Nallanathan,~\IEEEmembership{Fellow,~IEEE}
\thanks{Yuqin~Liu, Mona~Jaber, and Arumugam~Nallanathan are with the Queen Mary University of London, London E1 4NS, U.K. (e-mail: \{yuqin.liu, m.jaber, a.nallanathan\}@qmul.ac.uk).

Yan~Liu is with the College of Electronic and Information Engineering, Tongji University, Shanghai 201804, China. (e-mail: yanliu2022@tongji.edu.cn).
}
}

\maketitle
\begin{abstract}
A non-orthogonal multiple access (NOMA) assisted semi-grant-free (SGF) framework is proposed to enable channel access for grant-free users (GFUs) by using residual resources from grant-based users (GBUs). Under this framework, the problem of joint beamforming design and transmission scheduling is formulated to improve the system throughput and reduce the age-of-information (AoI) of GFUs. The aforementioned problem is transfer into a Markov Decision Process (MDP) to model the changing environment with the transmission/ waiting/ retransmission of GFUs. In an effort to solve the pertinent problem, firstly, a deep reinforcement learning (DRL) based transmission scheduling approach is proposed for determining the optimal transmission probability based on the available transmission slots and transmission status of GFUs. Secondly, a hierarchical learning algorithm is proposed to analyze the channel state information of GBUs and the transmission status of GFUs, and to train an upper-level policy based on this analysis for beamforming to achieve efficient grant-based (GB) transmission, while a lower-level policy adapts to maximize the utilization of transmission slots allocated by the upper-level agent. The two policies interact to improve channel access and avoid collisions. Numerical results reveal that 1) The DRL based transmission scheduling outperforms existing adaptive and state-dependent baselines in AoI reduction, where an average three-time-slots-earlier-transmission can be obtained compared to the state-dependent choice, and five time slots earlier can be achieved when comparing to the adaptive choice; 2) The hierarchical learning algorithm is able to achieve approximately a 31.82\% gain while maintaining the average AoI of GFUs within 1.5 time slots. 3) The effectiveness of the hierarchical learning scheme in NOMA-assisted SGF system is validated across scenarios with GFUs counts from 1-5 times of GBUs.
\end{abstract}

\begin{IEEEkeywords}
Age-of-information (AoI), deep reinforcement learning (DRL), semi-grant-free (SGF) transmission, non-orthogonal multiple access (NOMA). 
\end{IEEEkeywords}

\section{Introduction}
3GPP Release 19 (R19) advances the capabilities of the fifth generation (5G) networks with a focus on meeting the stringent demands of industrial and real-time applications, particularly by enhancing support for time-sensitive and ubiquitous communications~\cite{3gppR19}. Next-generation multiple access is envisioned not only to enhance traditional mobile broadband (eMBB) services but also to accommodate massive connectivity and diverse requirements of emerging Internet of Things (IoT) devices. Services such as ultra-massive machine-type communications (umMTC) and enhanced ultra-freliable low-latency communications (euRLLC) are key components in shaping future wireless networks. In this context, grant-free (GF) transmission emerges as a promising technique, allowing users to initiate data transmissions without waiting for prior scheduling or explicit resource allocation. By removing the overhead of centralized coordination, GF transmission is particularly suited for scenarios characterized by sporadic, massive transmissions~\cite{vaezi2022cellular}.

{However, the lack of prior scheduling in GF transmission can increase the risk of collisions and packet loss, making the age of information (AoI) a crucial performance metric for capturing the freshness of received data.} In particular, the AoI depicts the elapsed time from the data update is generated. This is especially important for supporting latency-sensitive services targeted by emerging communication paradigms, such as umMTC and euRLLC. Non-orthogonal multiple access (NOMA) has shown to be helpful in reducing AoI by enabling more transmission opportunities for grant-free users (GFUs)~\cite{ding2023unveiling}. The integration of NOMA and GF transmission was initially proposed in~\cite{choi2017noma} to enhance the capacity of uplink ALOHA system. Various NOMA-assisted GF schemes have been explored in~\cite{shahab2020grant}. By doing so, additional subchannels can be created by NOMA without any bandwidth expansion~\cite{choi2017noma}, while explicit scheduling are eliminated by GF technique, making this combination well-suited for massive connections in IoT scenarios. For instance, a semi-grant-free (SGF) scheme introduced in~\cite{ding2019simple} for scenarios with an excessive number of active users, where GFUs are allowed to share channels reserved by grant-based users (GBUs). In this case, the signal of GFUs becomes interference to the GBU with which it shares a channel. Thus, two contention control mechanisms were developed in~\cite{ding2019simple} to keep the interference effectively suppressed based on the decoding order of the GBU and the GFU. Moreover, the amount of supportable GFUs was investigated from the interference tolerance of GBUs in~\cite{ding2023new}. And the pre-configuration of signal-to-noise ratio (SNR) levels was investigated to facilitate the channel sharing among the multiple GFUs~\cite{ding2023noma}. 

Unlike GB transmission, where resources are pre-allocated, GFUs contend for channel access in a distributed manner, which may result in collisions and transmission interruptions. To address this challenge, a frameless ALOHA was introduced in~\cite{yuhao23}, restricting access to only nodes with non-zero age-gains in each frame. Similarly,~\cite{Moradian24} developed a dynamic frame-slotted ALOHA variant, which adapts frame lengths based on age-gain thresholds to achieve better performance. In addition to collision avoidance schemes,~\cite{ding2023impact} indicated that the transmission probability (TP) is a key factor for improving channel access and avoiding collisions and results in {a} significant impact on AoI performance. The aforementioned works provided the fundamental analysis to facilitate the system evolution. However, existing mathematic analyses are constrained to highly simplified implementation due to the complexity introduced by {evaluating the state transition probability of multiple GFUs}. 

Motivated by above, in this article, we investigate a NOMA-assisted SGF system and propose a hierarchical learning algorithm to jointly improve the AoI performance and the system throughput. {In particular, the channel states of GBUs is analyzed first to enable efficient beamforming design for GBUs, which in turn increases the number of available transmission slots for GFUs. Then, the transmission scheduling of GFUs is trained with the changing transmission status of users.} As the availability of transmission slots and the set of active GFUs evolve over time, the channel competition dynamics among GFUs can be modeled as a Markov Decision Process (MDP). Based on this, we explore the employment of reinforcement learning (RL) tools to capture the AoI and throughput performance in the MDP. In particular, the lower-level agent focuses on the AoI minimization {of} GFUs under available transmitting slots, whereas the upper-level agent focuses on efficient beamforming design, trying to create more transmitting slots for the lower-level agent. The two agents interact to enhance the system throughput by balancing channel access and transmission collisions. The main contributions of this article can be summarized as follows:
\begin{itemize}
    \item We formulate a NOMA-assisted SGF system to enable channel access for GFUs by residual resources {unused in} GB transmissions. Based on this model, a joint AoI reducing and throughput improvement problem is formulated by optimizing transmission scheduling and beamforming design.
    \item We conceive a DRL based approach to solve the MDP problem of channel competition among GFUs. In contrast to existing adaptive and state-dependent solutions, the proposed approach is capable of adapting with the dynamic environment, minimizing the AoI within three time slots. 
    \item We develop a hierarchical learning algorithm to solve the joint optimization of AoI and throughput performance. The algorithm achieves approximately a 31.82\% gain {over a DRL based approach} while maintaining information fresh.
    \item We demonstrate that the implementation of NOMA-assisted SGF system improves system capacity without bandwidth expansion. Additional, the hierarchical learning algorithm performs well  when GFU counts five times that of GBUs.
\end{itemize}

\section{RELATED WORKS}
\subsection{AoI for ALOHA-Like Random Access Schemes}
NOMA was first applied in~\cite{choi2017noma} to enhance the capacity of uplink ALOHA transmission. In recent years, NOMA has gained attention for its potential in reducing the AoI, particularly in achieving significant performance gains in the low {SNR} regime~\cite{ding2023unveiling}. This capability positions NOMA as a promising candidate for supporting {mMTC} in next-generation networks, sparking considerable research interest in this area. From the perspective of performance analysis, optimal access probabilities with dynamic channel conditions over multiple power levels is investigated when AoI exceeds a threshold~\cite{yong24}. Performance from imperfect transmission occasions are also analyzed, e.g. throughput from imperfect CSI and imperfect SIC~\cite{fan2023throughput} {and} AoI from noisy channels~\cite{zhou2022performance}. From the perspective of resource allocation, power allocation policies for minimized AoI on TDMA and NOMA schemes was designed  in~\cite{jixuan24} under unknown environments. Additionally,~\cite{jiao23} address the integration of constraints such as power allocation, network stability, and throughput requirements within resource allocation frameworks.

In the meantime, GF transmission extends this capability by eliminating the explicit scheduling before transmission, making it suitable for massive IoT deployments~\cite{vaezi2022cellular}. Various NOMA-based GF schemes have been explo in~\cite{shahab2020grant}. Note that ALOHA-like {random access} schemes may lead to significant collisions in physical mMTC systems, effectively rendering communication highly unreliable~\cite{zhou2022performance}. To address this challenge, a frameless ALOHA was introduced in~\cite{yuhao23}, restricting access to only nodes with non-zero age-gains in each frame. Similarly,~\cite{Moradian24} developed a dynamic frame-slotted ALOHA variant, which adapts frame lengths based on age-gain thresholds to achieve better performance.

In addition to collision avoidance schemes, GF transmission can be integrated with GB systems as a hybrid approach. In this work, we investigate the feasibility of leveraging the {unused} resources of GBUs to support occasional transmissions by GFUs, offering a potentially effective tool for AoI reduction.

\subsection{RL-based solutions for AoI reduction}
DRL gains its popularity in solving non-convex optimization problems by refining strategies through interactions in uncertain environments. In terms of AoI minimization, a distributed Q-learning approach is utilized at the device level to find the optimal sampling policy, while a device selection scheme at the BS monitors the dynamics of the physical process with minimal energy consumption~\cite{pereira2024reinforcenment}. {Similarly}, distributed DQN agents are deployed on IoT devices, with a mixing network implemented at the BS for coordination~\cite{Sihua22}. To address the overestimation of Q-values inherent in the original DQN, a double-DQN structure is proposed in~\cite{qingxi22}. However, in complex scenarios such as vehicle-to-everything (V2X) communications~\cite{Mlika22} and reconfigurable intelligent surface (RIS)-assisted networks~\cite{feng22}, DQN faces limitations due to the curse of dimensionality associated with mixed-integer nonlinear programming problems. To overcome this, the deep deterministic policy gradient (DDPG) algorithm is employed in~\cite{Mlika22} to efficiently learn continuous coverage and power control decisions for timely information exchange. Similarly, in~\cite{feng22}, DDPG addresses high-dimensional variables for joint optimization of the RIS phase-shift matrix and packet service time. In the meantime, a multiple agent DDPG is applied in~\cite{Ansarifard24} for AoI minimization in hierarchical aerial computing frameworks with uncertain CSI. For problems with higher complexity, researchers tend to not only enhance RL tools, but also decompose the problems into manageable subproblems. For example, the age of incorrect information (AoII) minimization problem in semantic-empowe downlink NOMA system with constraints including average/peak power constraint, network stability and freshness requirement is split into two steps in~\cite{hong24}, including the minimization of the upper bound of Lyapunov drift-plus-penalty and the resource allocation by proximal policy optimization (PPO). Besides this, authors in~\cite{long2024exploiting} design a hierarchical structure to optimize UAV's sensing scheduling and transmission control in outer-loop while updating UAVs' trajectories in the inner-loop. Another hierarchical framework presented in~\cite{congwei25} utilized two distributed agents to address scheduling efficiency and fairness, with a high-level agent balancing potential conflicts. Recently, integrated satellite-terrestrial networks (ISTNs) have emerged as key enablers for seamless global communication coverage, heightening the importance of maintaining information freshness for ubiquitous connections. To address these demands, a multi-agent PPO approach is proposed in~\cite{xinyu24}, employing decentralized execution for power allocation and interference cancellation, while centralized training is performed using generalized advantage estimation.

\subsection{Organization and Notations}
The rest of this paper is organized as follows. The NOMA-assisted SGF system is described in~\hyperref[NOMA-assisted SGF system]{Section~III}, and the problem formulation is illustrated in~\hyperref[Problem Formulation]{Section~IV}. \hyperref[Problem Formulation]{Section~V} elaborates on the proposed hierarchical learning algorithm for solving the formulated problem. The simulation results are detailed in~\hyperref[Problem Formulation]{Section~VI}. \hyperref[Conclusion]{Section~VII} concludes the main concept, insights, and the results of this work.

\section{NOMA Assisted Semi-Grant-Free Transmission}\label{NOMA-assisted SGF system}
\subsection{System description}
\begin{figure*}[ht]
    \begin{center}
        \centering
        \includegraphics[scale=0.45]{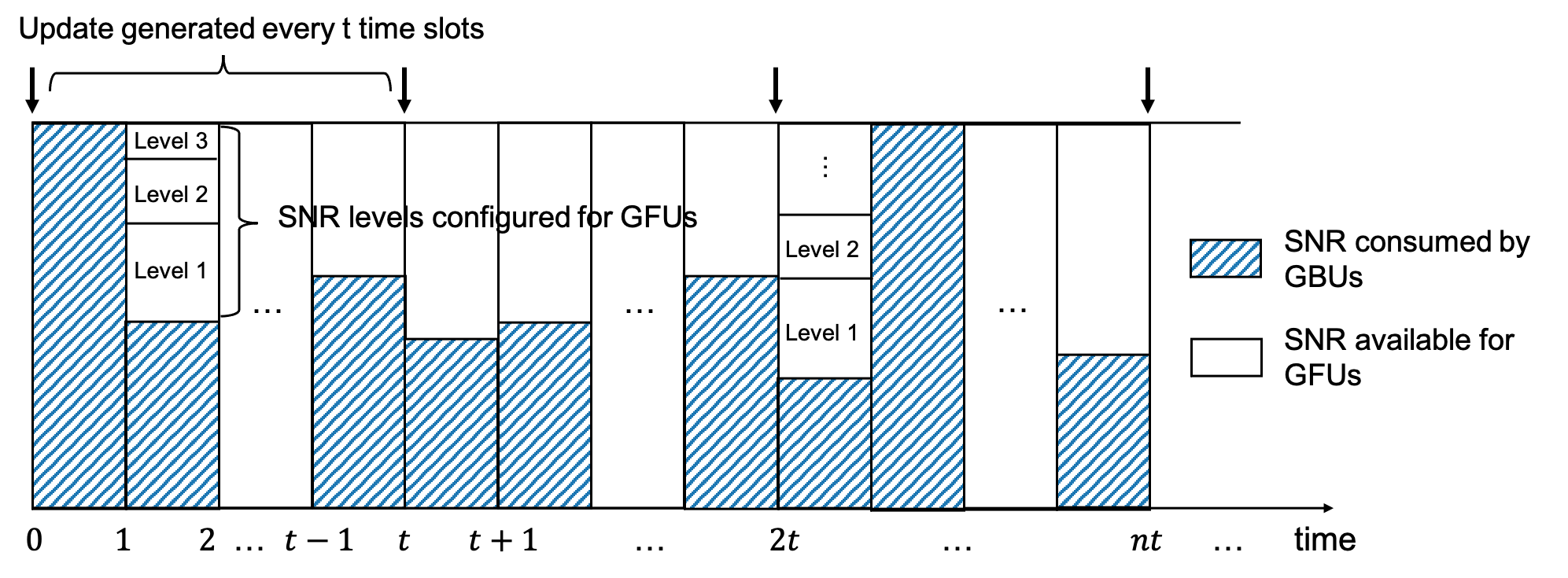}
        \caption{Considered SGF System}
        \label{system model}
    \end{center}
\end{figure*}

We consider an uplink grant-based transmission, where there are $K$ GBUs connected to the BS. In this work, the key idea of the SGF framework is to achieve extra throughput by providing additional links to GFUs when there are remaining resources from GBUs, as well as creating more transmitting opportunities for GFUs. {The considered SGF system is presented in Fig.~\ref{system model}.}

To this end, GFUs are allowed to access to the network following the principle of random access NOMA~\cite{choi2017noma}. Meanwhile, a Generate-at-Request (GAR) policy is applied so that each GFU can generate an update every few time slots, and the GFU will continue to retransmit until it is successful~\cite{ding2023age}. We also assume that GFUs always have a message to transmit and that GFUs who have successfully transmitted their updates will not compete for slots until the next update is generated. Also, we assume GFUs are able to receive information of the outcome of their updates via a dedicated control channel. The BS aims to make the messages from users as fresh as possible with respect to the AoI by optimizing control variables.

\subsection{Interference tolerance}
Since GF transmissions are enabled only if the experience of GBUs is guaranteed, we prioritize GBUs' transmission in this system. In this case, the BS will decode GBUs' signal first, and GFUs will post an interference to the GBU they are admitted to. Therefore, it is necessary to know how much interference can be tolerated by each GBU, as {this value} determines how many GFUs can be supported. Given this, the maximum interference {that could be tolerated} by the GBU $k$ is denoted by
\begin{align}
    I_k^{max} = \frac{P_k {|\bm{h}_k \bm{v}_k|}^2}{2^{\hat{R}_k} - 1} - n_0,
\end{align}
where $P_k$ is the transmit power of the GBU $k$, $\bm{h}_k$ is the channel info of the GBU $k$, $\bm{v}_k$ is the detecting vector of the GBU $k$, $\hat{R}_k$ is the expected rate of GBU $k$, and {$n_0$ is the additive white Gaussian noise (AWGN)}.

Then, the interference budget for GFUs is given by
\begin{align}
    I_k^{GF} = I_k^{max} - I_k^{GB}, \label{I_GF}
\end{align}
where $I_k^{GB}$ is the interference from other GBUs, and
\begin{align}
    I_k^{GB} = \sum_{j\neq k}P_j |\bm{h}_k \bm{v}_j|^2 \label{I from GBU}.
\end{align}

\subsection{Pre-confirguration of SNR levels}
Given the prioritized GB transmission, the BS will pre-configure SNR levels for GFUs to access when there are leftovers from GBUs. In this work, we would follow a particular design of the SNR pre-configuration, which was originally proposed in~\cite{ding2023noma}. In particular, SNR levels from GBU $k$ in this work, denoted by $\Gamma_k^n$, can be configu as
\begin{align}
    \Gamma_n^N = 2^{{\hat{R}_k}} - 1, (1 \leq n \leq N),& \\
    \Gamma_k^n = (2^{{\hat{R}_k}} - 1)(1 + \Gamma_k^{n+1}), \label{SNRlevels}
\end{align}
where $N$ is the accepted number of GFUs, and $N$ needs to satisfy the constraint
\begin{align}
    \sum_{n=1}^{N}\Gamma_k^N \leq I_k^{GF}. \label{SNRconstraint}
\end{align}

For an occasion with $N >0$, GFUs who haven't sent their update to the BS will make a transmission attempt with transmission probability $\mathbb{P}_{\text{TP}}$. Assume that the GFU$_{k'}$ has successfully secured one of the SNR levels $\Gamma_k^n$ with probability $\mathbb{P}_{\text{TP}}$, the user will scale its transmitted signal by $\frac{\Gamma_k^n n_0}{{|\bm{h}_{k'} \bm{v}_k|}^2}$. To guarantee the chosen SNR level is feasible for the successive interference cancellation (SIC), it is assumed that
\begin{align}
    \frac{\Gamma_k^n n_0}{{|\bm{h}_{k'} \bm{v}_k|}^2} \leq \Gamma,\label{SICok}
\end{align}
where $\Gamma$ denotes the GFU's transmit SNR.

\subsection{AoI model}
With the aim to enhance the data freshness of GFUs, the concept of AoI is introduced to quantify the data freshness performance~\cite{kaul2012real}. The AoI of an update is the duration from when it is generated to the current time. Specifically, the AoI of an update will increase with time until it is transmitted. And the AoI will drop to the duration of one time slot after transmission as we assume the update will at least consumes one time slot to be sent to the receiver. 

Based on above, the AoI of GFU $k'$ is updated by
\begin{align}\label{AoIupdate}
\Delta_{k'}[t] =
\left\{
\begin{array}{ll}
1, & \text{transmission successful,} \\
\Delta_{k'}[t-1]+1, & \text{transmission failed,}
\end{array}
\right.
\end{align}
where the instantaneous AoI of the GFU $k'$ at time $t$ is calculated by~\cite{sun2017update}:
\begin{equation}
    \Delta_{k'}[t] = t - T_{k'},
\end{equation}
where $T_{k'}$ is the generation time of the freshest update successfully delivered from the GFU $k'$. Note GFUs will fail to transmit {for any of the following reasons}:
\begin{itemize}
    \item There is no remaining resource from GBUs;
    \item The GFU does not make an attempt for transmission;
    \item The chosen SNR level is not feasible for SIC, i.e., $\frac{\Gamma_k^n n_0}{{|h_{k'} v_k|}^2} > \Gamma$;
    \item There are more transmitted updates than available slots, which leads to a collision for GFUs sharing the same resource block.
\end{itemize}

Thus the minimization of AoI {entails} finding a trad-off between increasing channel access and avoiding collisions.

\section{Problem Formulation}\label{Problem Formulation}
We aim to reduce the AoI of GFUs and increase the system throughput, which is attained by a jointly optimization of transmission scheduling and beamforming design. Since the available slots for GFUs' channel competition depend on the residual resources from GB transmission, the original problem can be decoupled into two sequential subproblems {as detailed in the following sections.}






\subsection{AoI Minimizing for GFUs}
The average AoI of {all $K'$} GFUs at time slot $t$ is denoted by
\begin{align}
    \Bar{\Delta} = \frac{1}{K'} \sum_{k'=1}^{K'} \sum_{t=0}^{T}\Delta_{k'}[t].
\end{align}

With the aim to enhance the data freshness by the idle SNR levels, the transmission probability, $\mathbb{P}_{\text{TP}}$, is optimized for increasing channel access and reducing collisions. Thus the AoI minimizing problem is written as
\begin{align}\label{P1}
\mathcal{P}_1: &\min_{\mathbb{P}_{\text{TP}}} \Bar{\Delta}\\
\textrm{s.t.}
&\sum_{n=1}^{N}\Gamma_k^N \leq I_k^{GF}, \forall k\in K, \forall n \in N, \label{c1} \\ 
&\frac{\Gamma_k^n n_0}{{|\bm{h}_{k'} \bm{v}_k|}^2} \leq \Gamma, \forall k\in K, k' \in K', \forall n \in N, \label{c2}
\end{align}
where constraint~\eqref{c1} is the interference tolerance of GBUs, and constraint~\eqref{c2} is the SNR constraint for SIC process.

\subsection{Throughput Maximization for SGF NOMA Systems}
The throughput of the system comes from the transmission of the GBUs and GFUs, where the throughput of GBU $k$ and GFU $k'$ at time slot $t$ are denoted by
\begin{align}
    R_k[t] = \log_{2}({1 +\frac{P_k {|\bm{h}_k \bm{v}_k|}^2}{I_k^{GB} + I_k^{GF} + n_0}}),\\
    R_{k'}[t] = \log_{2}({1 +\frac{P_{k'} {|\bm{h}_{k'} \bm{v}_k|}^2}{I_{k'}^{GF} + n_0}}).
\end{align}

{It is expected that additional transmit opportunities are likely to result in an increase in throughput. Including beamforming tuning in the joint optimization problem is expected to improve the efficiency of GBUs transmission and thus to provide more opportunities for GBUs to transmit.} Thus the {throughput improvement} is achieved by the optimized beamforming and organized transmission scheduling. The throughput of the NOMA-assisted SGF system in duration $T$ is given by
\begin{subequations}
\begin{align}\label{P2}
\mathcal{P}_2: &\max_{\bm{V}}({\sum\limits_{k=1}^{K}\sum_{t=0}^{T}R_k[t] + \sum\limits_{k'=1}^{K'}\sum_{t=0}^{T}R_{k'}[t]})\\
\textrm{s.t.}
&R_k[t] \geq \hat{R}, \forall k \in K, \label{s1}
\end{align}
\end{subequations}
where constraint~\eqref{s1} is the constraint that GBU's transmission is guaranteed.

\section{Proposed Solution}\label{proposed slution}
\subsection{Markov Decision Process Formulation}
{As explained earlier, a GFU that successfully transmitted will not reattempt transmission until a new update is generated. In other words, the transmission status in the current time slot will influence the number of transmitting GFUs.}
Thus, the channel competition among GFUs can be formulated as a MDP. In each MDP cycle, the process carries out actions according to the current state and the outcome from the last cycle, and then transits into a new state.



\subsection{DRL for AoI Minimizing}\label{RL4AoI}
\begin{figure}[t]
    \centering
    \begin{subfigure}{\linewidth}
        \centering
        \includegraphics[width=0.95\linewidth]{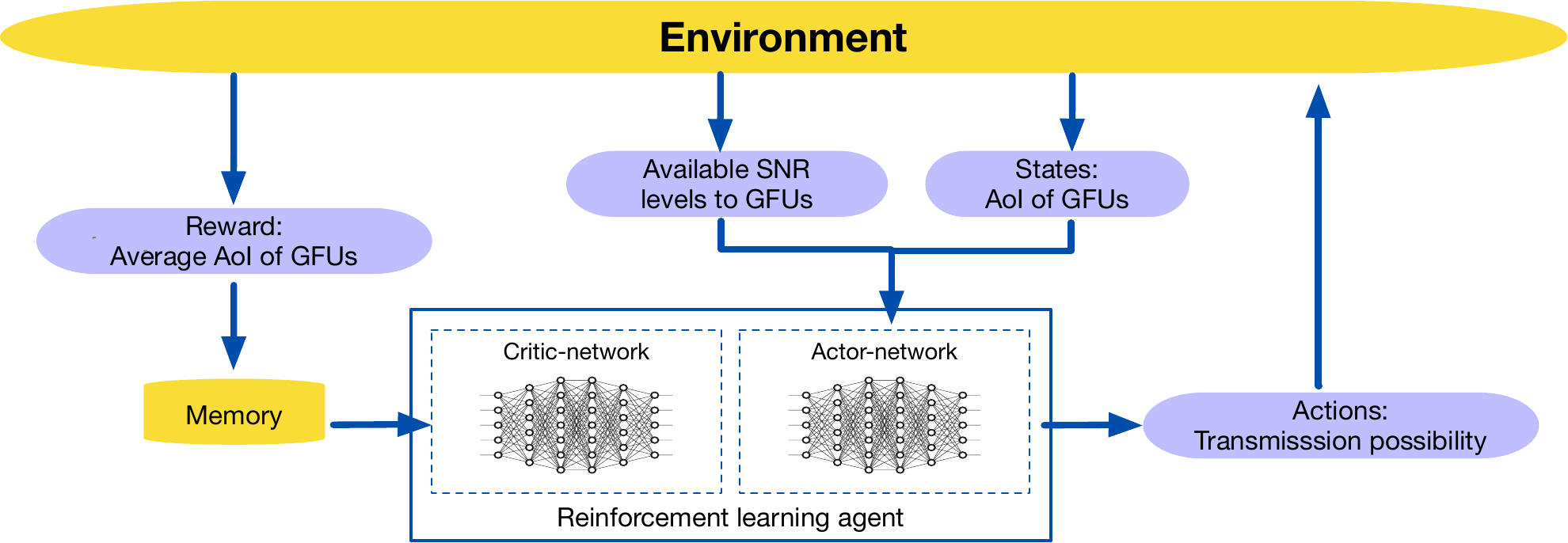}
        \caption{Deep reinforcement learning}
        \label{RL_agent}
    \end{subfigure}
    \begin{subfigure}{\linewidth}
        \centering
        \includegraphics[width=0.95\linewidth]{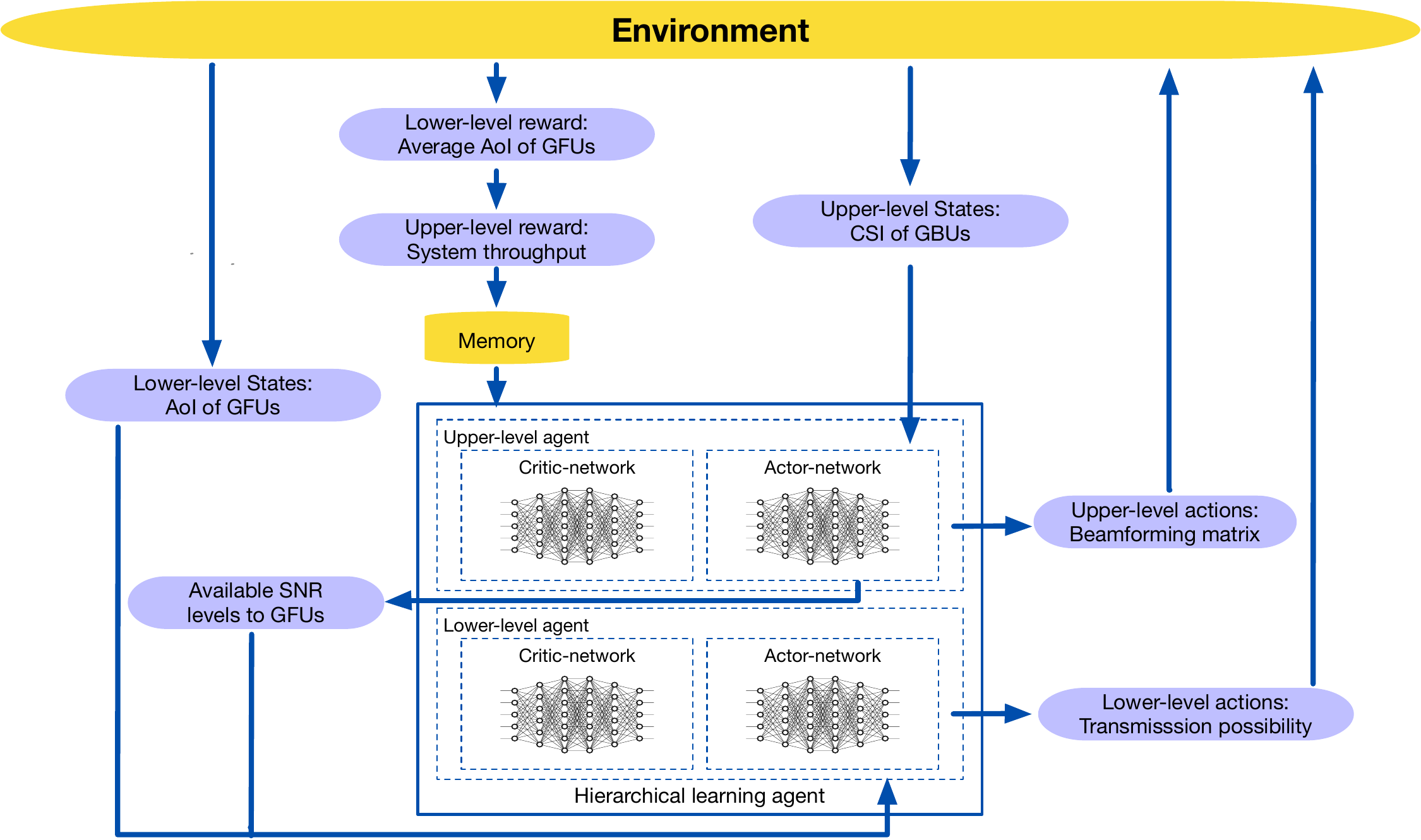}
        \caption{Hierarchical learning.}
        \label{HL_agent}
    \end{subfigure}
    \caption{Hierarchical learning solution for the formulated problem.}
    \label{HL_flow}
\end{figure}

In this section, a DRL based solution is introduced to optimize TP, $\mathbb{P}_{\text{TP}}$, for AoI minimizing. A fixed choice, a state-dependent choice~\cite{bae2022age}, and an adaptive choice~\cite{ding2023impact} {were previously proposed.} And we would like to explore if DRL can be helpful for this problem, {a flow diagram of the DRL base solution can be found in Fig.~\ref{HL_flow}(\subref{RL_agent})}.

We aim to minimize the average AoI by optimizing the TP, $\mathbb{P}_{\text{TP}}$, of GFUs. The state space, action space, and rewards for this MDP process are defined as follows:

\textbf{State space}: The instantaneous AoI of each GFU are modeled as states, denoted by
\begin{align}
    \bm{s}_l[t] = \{ \Delta_1[t], \Delta_2[t],... , \Delta_{k'}[t], \forall k'\in K'\},
\end{align}
where the AoI of each GFU is updated by Eq.~\eqref{AoIupdate}.


\textbf{Action space}: An identical TP, $\mathbb{P}_{\text{TP}}$, for all waiting GFUs is modeled as the action, denoted by
\begin{align}
    \bm{a}_l[t] = \{ \mathbb{P}_{\text{TP}}[t], \mathbb{P}_{\text{TP}} \in [0, 1] \}.
\end{align}

A large $\mathbb{P}_{\text{TP}}$ is expected to avoid idle resources and encourage more transmissions, whereas a smaller $\mathbb{P}_{\text{TP}}$ is expected to do the opposite to avoid collisions. Also, the GFU that transmitted successfully will keep silent until the next update generated. Thus there tend to be less GFUs in the competition in each generation period. The DRL agent aim to balance the available resource and competing GFUs by optimizing $\mathbb{P}_{\text{TP}}$.

\textbf{Rewards}: The optimization target for this subproblem is to minimize the AoI of GFUs. Hence the reward function is formulated by the average AoI of GFUs, denoted by
\begin{align}\label{r_inner}
    \bm{r}_l[t] = \frac{1}{K'} \sum_{k'=1}^{K'} \Delta_{k'}[t].
\end{align}

During the learning, {the policy $\pi_\theta(a_l[t]|s_l[t])$ defines the action to be taken in each state $s_l[t]$, and the agent aims to find the optimal policy $\pi^*_\theta$ by minimizing a clipped surrogate loss function that encourages improvement while constraining deviation from the current policy.}
\begin{align}
    L^{\text{CLIP}}(\theta) = \hat{\mathbb{E}}_t[\min (r_l[t](\theta)\hat{A}[t], clip(r_l[t](\theta), 1-\epsilon, 1+\epsilon)\hat{A}[t])], \label{policy}
\end{align}
where $r_l[t](\theta) = \frac{\pi_\theta(a_l[t]|s_l[t])}{\pi_{\theta_{\text{old}}}(a_l[t]|s_l[t])}$ is the probability ratio, $\epsilon$ is the hyperparameter to {limit how much the new policy can deviate from the old policy, $\hat{A}[t]$ is the estimated advantage at timeslot $t$, denoted by}
\begin{align}
    \hat{A}[t] = r_l[t] + \gamma V_\pi(s_{l}[t+1]) - V_\pi(s_l[t]),
\end{align}
where $V_\pi$ is a learned state-value function of policy $\pi$. {The training process is summarized in Algorithm~\ref{ppo}.}

\subsection{Hierarchical Learning for Joint AoI Minimizing and Throughput Improvement}
In this part, we extend the investigation of transmission scheduling to the joint optimization of transmission scheduling and beamforming design. The reason of this design is that {additional transmit opportunities obtained from the optimized transmission scheduling are likely to result in an increase in throughput. Including beamforming tuning in the joint optimization problem is expected
to improve the efficiency of GBUs transmission and thus to provide more opportunities for GBUs to transmit. Thus a joint optimization of transmission scheduling and beamforming can contribute for throughput improvement, as well as AoI performance. As the amount of available transmitting slots is up to the amount of residual resource from GB transmissions, the joint optimization is decoupled into two sequential sub-tasks, where a hierarchical learning approach is proposed as shown in Fig.~\ref{HL_flow}(\subref{HL_agent}).}

{The first step of the hierarchical learning is to achieve an efficient beamforming for GFUs so as to enable more transmitting opportunities for the waiting GFUs. Thus} the optimization of beamforming design is processed at the upper-level, where GBUs' CSI is taken as the state, denoted by
\begin{align}
    \bm{s}_u[t] = \{\bm{h}_1[t], ..., \bm{h}_k[t], \forall{k} \in K\}.
\end{align}

{And the action space is the detecting matrix of GBUs, which is noted as:}
\begin{align}
    \bm{a}_u[t] = \{ \bm{V}[t] \} 
\end{align}

The accepted interference from GFUs is defined in~\eqref{I_GF}, and the number of SNR levels as defined in~\eqref{SNRlevels}. Given the available slots from the upper-level, the transmission scheduling will be optimized by the agent designed in the Section~\ref{RL4AoI}. 

After that, the reward function is expressed as
\begin{align}
\bm{r}_u[t] = \sum\limits_{k=1}^{K}R_k + \sum\limits_{k'=1}^{K'}R_{k'}.\label{upper_reward}
\end{align}

The policy of the upper-level is adapted by the PPO as well. While different update frequencies are set due to the different sensitivities of throughput and AoI performance to time.

During the learning, both the upper-level policy $\pi_\zeta(a_u|s_u)$ and the lower-level policy $\pi_\theta(a|s)$ are updated by~\eqref{policy}.

Although {both policies follow} the same policy gradient strategy, different update frequencies are set for $\pi_\zeta$ and $\pi_\theta$ due to the different sensitivities of throughput and AoI performance to time. Then, with the update frequencies for each policy, collected transitions are evaluated to update the policy by the empirical average advantage at the terminal of each sample batch:
\begin{align}\label{closs}
    L^{\mathrm{critic}} = \frac{\sum\limits_{t=0}^{T} \sqrt{\hat{A}[t]}}{t},
\end{align}
where $L^{\mathrm{critic}}$ is the loss of critic network. {The training is summarized in Algorithm~\ref{HRL}.}

\begin{algorithm}[t]
\caption{DRL for AoI Minimizing \eqref{P1}}
\label{ppo}
\begin{algorithmic}[1]
        \STATE Initialize the environment: the AoI of GFUs $\Delta_{k'}$, the generation time of GFUs' update $T_{k'}$, the update generation frequency $f$.
        \STATE Initialize the actor-network, the critic-network, and the sample batch.
        \FOR{each episode}
        \STATE Reset the environment
        \FOR{each step $t$}
        \IF {$t \% f = 0$}
            \STATE Update the update generation time: $T_{k'} \leftarrow t$
            \STATE Update the number of GFUs: $\text{waiting GFU} \leftarrow K'$
        \ENDIF
        \STATE Observe the current state $\bm{s}[t] = \{ \Delta_1[t], \Delta_2[t],... , \Delta_{k'}[t], \forall k'\in K'\}$
        \STATE Generate an action $\bm{a}[t] = \{ \mathbb{P}_{\text{TP}}[t], \mathbb{P}_{\text{TP}} \in [0, 1] \}$ by policy $\pi_\theta(\bm{a}|\bm{s})$
        \FOR{each waiting GFU $k'$}
        \STATE Carry out a transmit attempt with $\mathbb{P}_{\text{TP}}[t]$
        \STATE \textbf{case} Transmission successful: $\Delta_{k'} = 1$, $\text{waiting GFU} \leftarrow \text{waiting GFU} - 1$
        \STATE \textbf{case} Transmission failed: $\Delta_{k'} = t + 1$
        \ENDFOR
        \STATE Calculate the reward $\bm{r}[t]$ by~\eqref{r_inner}
        \STATE Save transition $<\bm{s}[t], \bm{a}[t], \bm{r}[t], \bm{s}[t+1]>$ into the sample batch
        \IF {$t$ $\%$ replace iter $=$ 0}
            \STATE update the policy $\pi_\theta$ by~\eqref{policy}
        \ENDIF
        \ENDFOR
        \ENDFOR
\end{algorithmic}
\end{algorithm}

\begin{algorithm}[t]
\caption{Hierarchical Learning for Joint AoI Reducing and Throughput Improvement~\eqref{P2}}
\label{HRL}
\begin{algorithmic}[1]
        \STATE Initialize the environment: the channel of GBUs $\bm{h}_k$, the channel of GFUs $\bm{h}_{k'}$, AoI of GFUs $\Delta_{k'}$, generation time of GFUs' update $T_{k'}$, the update generation frequency $f$.
        \STATE Initialize the hierarchical learning agent: the upper-level policy $\pi_\zeta$, the lower-level policy $\pi_\theta$.
        \FOR{each episode}
        \STATE Reset the environment
        \FOR{each step $t$}
		\STATE Take an observation $\bm{s}_u[t] = \left\{\bm{h}_1[t], ..., \bm{h}_k[t], \forall{k} \in K\right\}$ from the environment
            \STATE Generate an action $\bm{a}_u[t] = \{ \bm{V}[t] \}$ by the upper-level policy $\pi_\zeta$
            \STATE Calculate the interference tolerance of GBUs by~\eqref{I_GF}
            \STATE Divide the SNR levels for GFUs by~\eqref{SNRlevels}
            \IF {$t \% f = 0$}
            \STATE Update the update generation time: $T_{k'} \leftarrow t$
            \STATE Update the number of GFUs: $\text{waiting GFU} \leftarrow K'$
        \ENDIF
        \STATE Observe the current state $\bm{s}_l[t] = \{ \Delta_1[t], \Delta_2[t],... , \Delta_{k'}[t], \forall k'\in K'\}$
        \STATE Generate an action $\bm{a}_l[t] = \{ \mathbb{P}_{\text{TP}}[t], \mathbb{P}_{\text{TP}} \in [0, 1] \}$ by policy $\pi_\theta(\bm{a}_l|\bm{s}_l)$
        \FOR{each waiting GFU $k'$}
        \STATE Carry out a transmit attempt with $\mathbb{P}_{\text{TP}}[t]$
        \STATE \textbf{case} Transmission successful: $\Delta_{k'} = 1$, $\text{waiting GFU} \leftarrow \text{waiting GFU} - 1$
        \STATE \textbf{case} Transmission failed: $\Delta_{k'} = t + 1$
        \ENDFOR
        \STATE Calculate the lower-level reward $\bm{r}_l[t]$ by~\eqref{r_inner}
        \STATE Calculate the upper-level reward $\bm{r}_u[t]$ by~\eqref{upper_reward}
        \STATE Save transition $<\bm{s}_l[t], \bm{a}_l[t], \bm{r}_l[t], \bm{s}_l[t+1]>$ into the sample batch
        \STATE Save transition $<\bm{s}_u[t], \bm{a}_u[t], \bm{r}_u[t], \bm{s}_u[t+1]>$ into sample batch
        \IF {$t \%$ upper-level parameter replace iter $=$ 0}
            \STATE update the policy $\pi_\theta$
        \ENDIF
        \IF {$t \%$ lower-level parameter replace iter $=$ 0}
            \STATE update the policy $\pi_\zeta$
        \ENDIF
        \ENDFOR
        \ENDFOR
\end{algorithmic}
\end{algorithm}







\section{Complexity Analysis}
The joint optimization of the hierarchical learning follows the policy adaptation of the PPO, where the upper-level agent and the lower-level agent are iterated separately in frequency $f_\theta$ and $f_\zeta$. As suggested in~\cite{Samir22}, the complexity of PPO is denoted by $O(\sum\limits_{j=1}^{J}n_{j-1}n_{j})$, where $n_j$ is the number of neurons in layer $j$ among $J$ layers. As the upper-level policy and the lower-level policy are updated separately, the overall complexity of the hierarchical learning can be expressed by $O(f_\theta \sum\limits_{i=1}^{I}n^u_{i-1}n^u_{i} + f_\zeta \sum\limits_{j=1}^{J}n^l_{j-1}n^l_{j})$, where the $n^u_i$ and $n^l_j$ are the number of neurons of the upper-level agent and the lower-level agent.

\section{Simulation Results}
In this section, we provide the numerical results of the DRL and the hierarchical learning algorithm for solving joint AoI and throughput optimization in NOMA-assisted SGF systems. In this simulation, we consider that GBUs and GFUs are independently distributed following a normal distribution cente at the BS, with each coordinate dimension having zero mean and a standard deviation of 1.5 km, and users are roaming randomly at each time slot. The other simulation parameters are listed in Table~\ref{tab:my_label} unless otherwise stated.

\begin{table}[t]
    \caption{Simulation parameters}
    \centering
    \setlength{\tabcolsep}{0.45cm}{
    \begin{tabular}{|c|c|c|}
    \hline
         Parameter & Description & Value  \\ \hline
         $N$ & number of BS antennas & 3  \\ \hline
         $K$ & number of GBUs & 3  \\ \hline
         $K'$ & number of GFUs & 5  \\ \hline
         $f$ & update generation frequency & 3  \\ \hline
         $P$ & GBUs' transmit power & 23 dBm  \\ \hline
         $B$ & bandwidth & 1 MHz  \\ \hline
           $R$ & Rician factors & 1 \\ \hline
         $\sigma$ & white Gaussian noise & -110 dBm  \\ \hline
         $\gamma$ & discount factor & 0.99 \\ \hline
         $lr_{actor}$ & learning rate of actor-network & 0.0003 \\ \hline
         $lr_{critic}$ & learning rate of critic-network & 0.001 \\ \hline
         $\epsilon$ & clip ratio & 0.1  \\ \hline
         $e$ & number of epochs & 5  \\ \hline
    \end{tabular}}
    \label{tab:my_label}
\end{table}

\subsection{DRL for AoI Minimizing}
In this section, we present the performance of DRL based solution for solving the AoI minimizing problem, as formulated in~\eqref{P1}. As the baseline models, two existing settings of TP $\mathbb{P}_{\text{TP}}$ are implemented and compared with the proposed solution:
\begin{itemize}
    \item Adaptive choice~\cite{ding2023impact}: $\mathbb{P}_{\text{TP}} = \min\{1, \frac{L}{M}\}$, 
    \item State-dependent choice~\cite{bae2022age}: $\mathbb{P}_{\text{TP}} = \frac{1}{M-j}$,
\end{itemize}
where {$L$} here is the number of SNR levels, $M$ is the number of GFUs, and $j$ is the GFUs that have transmitted successfully.

\subsubsection{Performance of DRL}
\begin{figure}[t]
    \centering
    \includegraphics[width=0.95\linewidth]{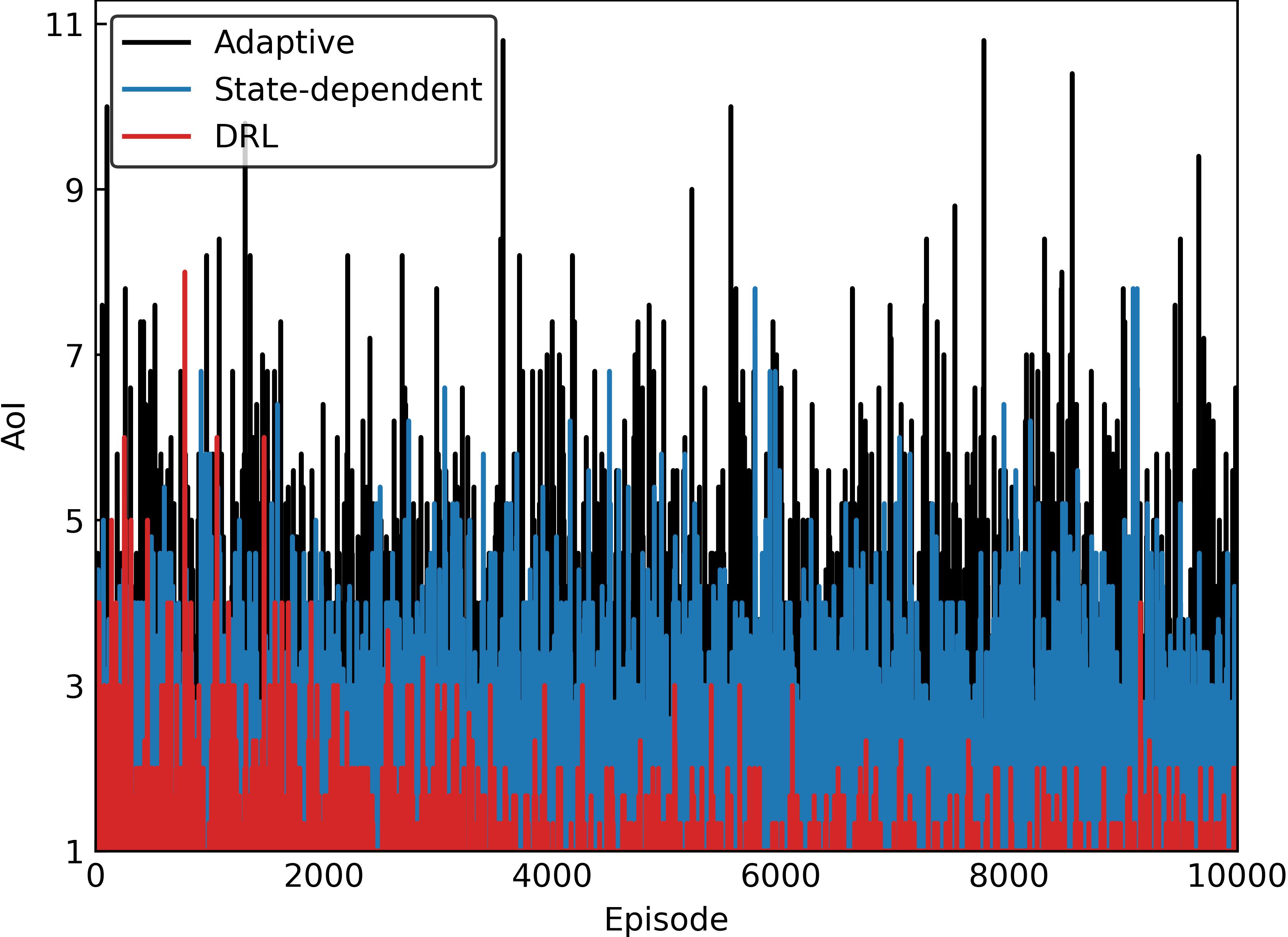}
    \caption{AoI performance.}
    \label{lower-level-AoI}
\end{figure}

Firstly, Fig.~\ref{lower-level-AoI} depicts the convergence of the DRL based approach, where the average AoI of the GFUs can be minimized to less than three time slots after 6000 episodes. Secondly, the adaptive choice~\cite{ding2023impact} and the state-dependent choice~\cite{bae2022age} were proposed to enable a dynamic setting of $\mathbb{P}_{\text{TP}}$ with decreasing waiting GFUs and regular update generation. It can be observed from Fig.~\ref{lower-level-AoI} that the DRL based approach is more effective in capturing the relation among the dynamic issues and delivering a better response, where approximately three time slots gain can be obtained for each GFU compa to the state-dependent choice, and five time slots gain compa to the adaptive choice.

\subsubsection{Throughput Gain from AoI Minimizing}
\begin{figure}[t]
    \centering
    \includegraphics[width=0.95\linewidth]{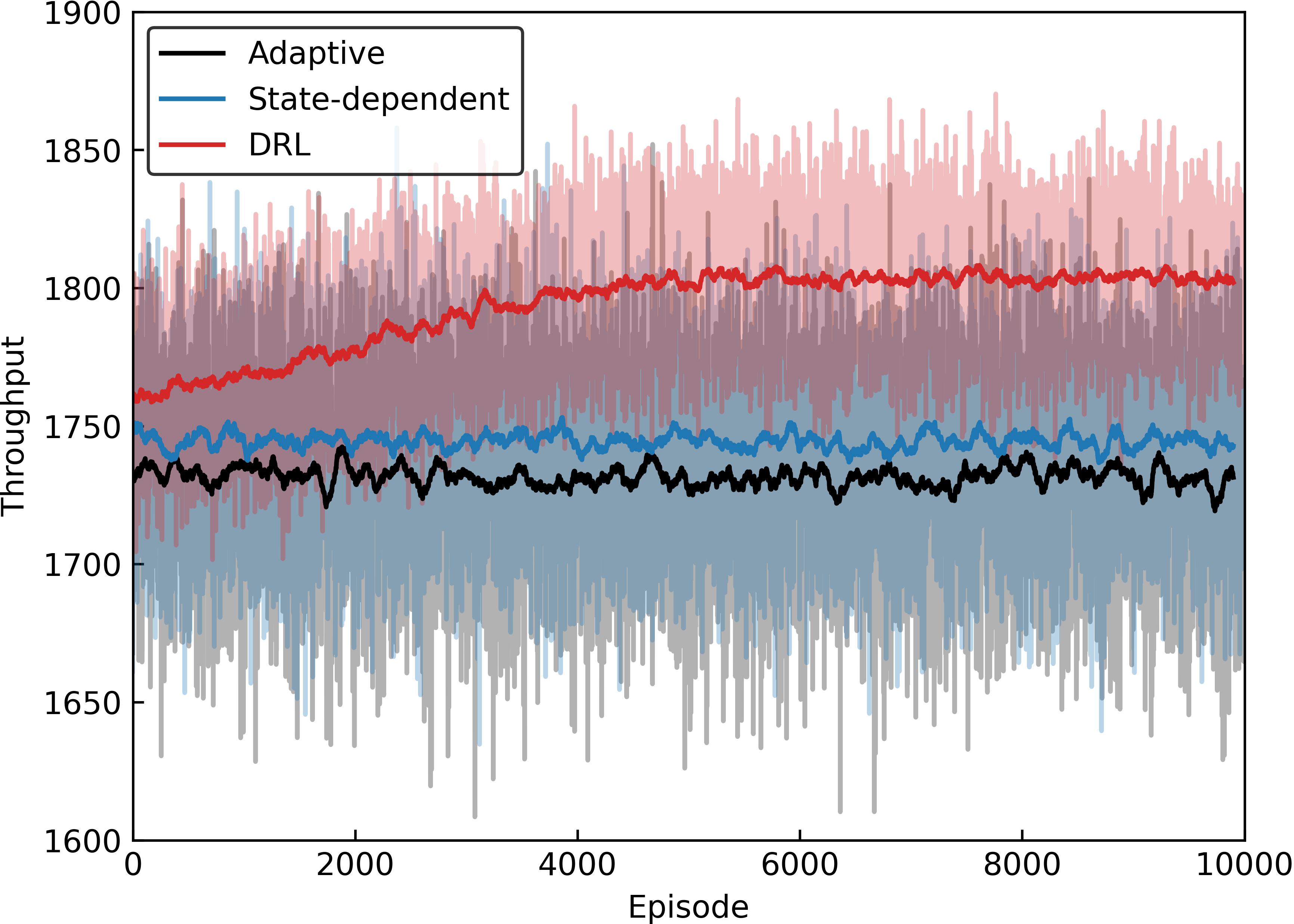}
    \caption{Throughput gain from AoI minimizing.}
    \label{lober-level-throughput}
\end{figure}

The AoI minimizing is helping to get GFUs transmitted earlier from the user's perspective. From the point of view of the network system, AoI minimizing is helping to bring more transmission in each time slot. In this way, higher throughput can also be obtained by optimizing AoI. To verify this thinking, we print the throughput from the three AoI optimizing schemes. As shown in Fig.~\ref{lober-level-throughput}, additional 50-70 bit/s of throughput can be obtained from AoI minimizing by DRL approach compa to the two baselines.

\subsection{Hierarchical Learning for Joint AoI Minimizing and Throughput Improvement}
In this section, we demonstrate the simulation results of the hierarchical learning algorithm when applied to the joint optimization of AoI and throughput, as formulated in~\eqref{P1} and~\eqref{P2}.
\subsubsection{Overall Performance of Hierarchical Learning}
\begin{figure}[t]
    \centering
    \includegraphics[width=0.95\linewidth]{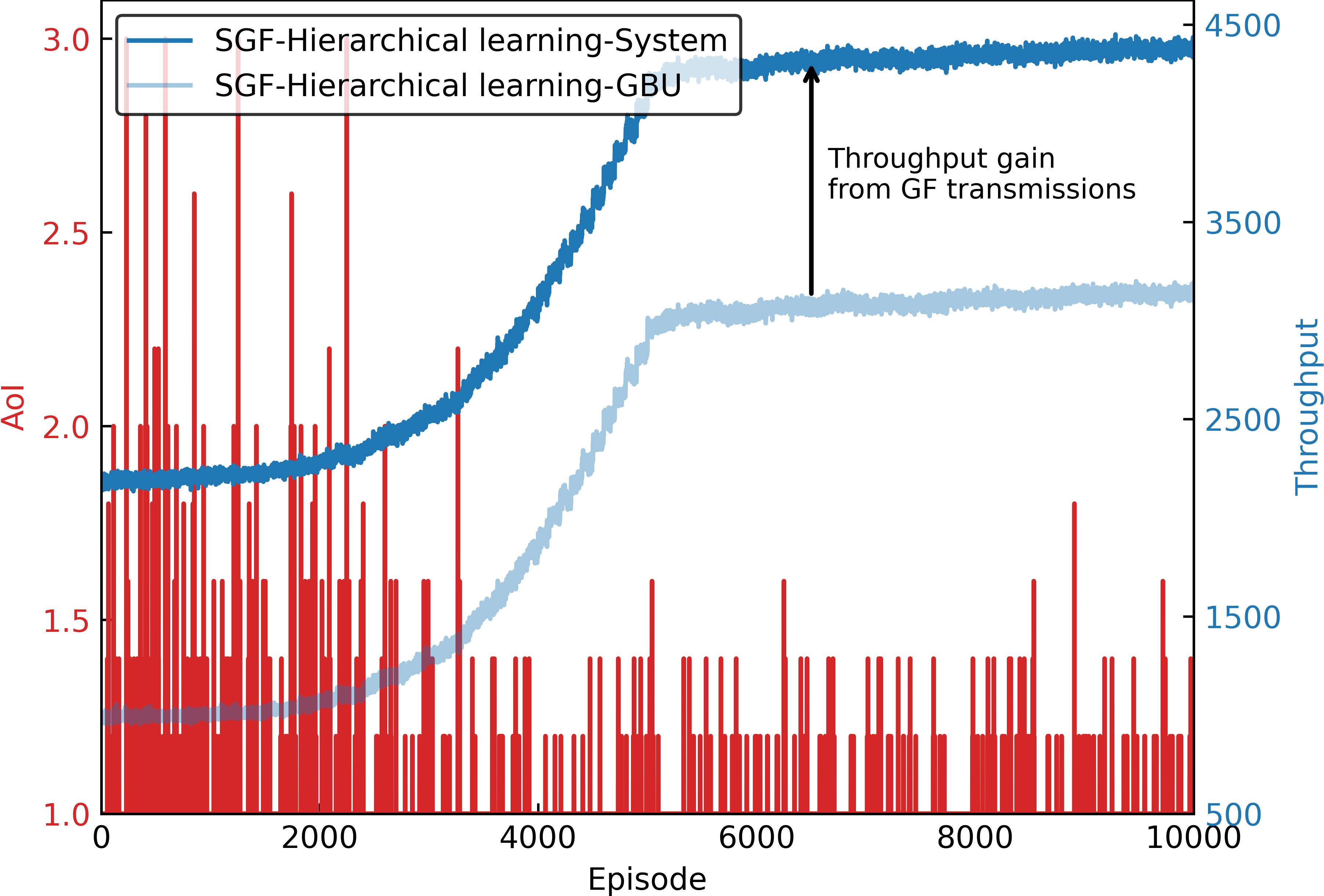}
    \caption{Upper-level rewards and lower-level rewards of the hierarchical learning algorithm versus the number of training episodes.}
    \label{SGF&AoI}
\end{figure}

Fig.~\ref{SGF&AoI} demonstrates the performance of the proposed algorithm, where the upper-level policy is adapted to fulfill GBUs’ transmission efficiently, and the lower-level policy is adapted to schedule GFUs' transmission for improving channel access and avoiding collisions. In Fig.~\ref{SGF&AoI}, the AoI performance mainly comes from the lower-level agent by adapting the TP $\mathbb{P}_{\text{TP}}$ with the available transmitting slot and number of waiting GFUs. While the throughout performance comes from (1) the transmission of GBUs, and (2) the additional transmissions enabled by the optimized $\mathbb{P}_{\text{TP}}$. The gap between the two blue lines illustrate the amount of transmissions from GFUs, which is in line with our motivation that serving GFUs by leftovers from GBUs is able to improve the system capacity. And, by the optimized scheduling policy, the GFUs are able to transmit their update within approximately 1.5 time slots.

\begin{figure*}[t]
    \centering
    \begin{subfigure}{0.45\linewidth}
        \includegraphics[width=\linewidth]{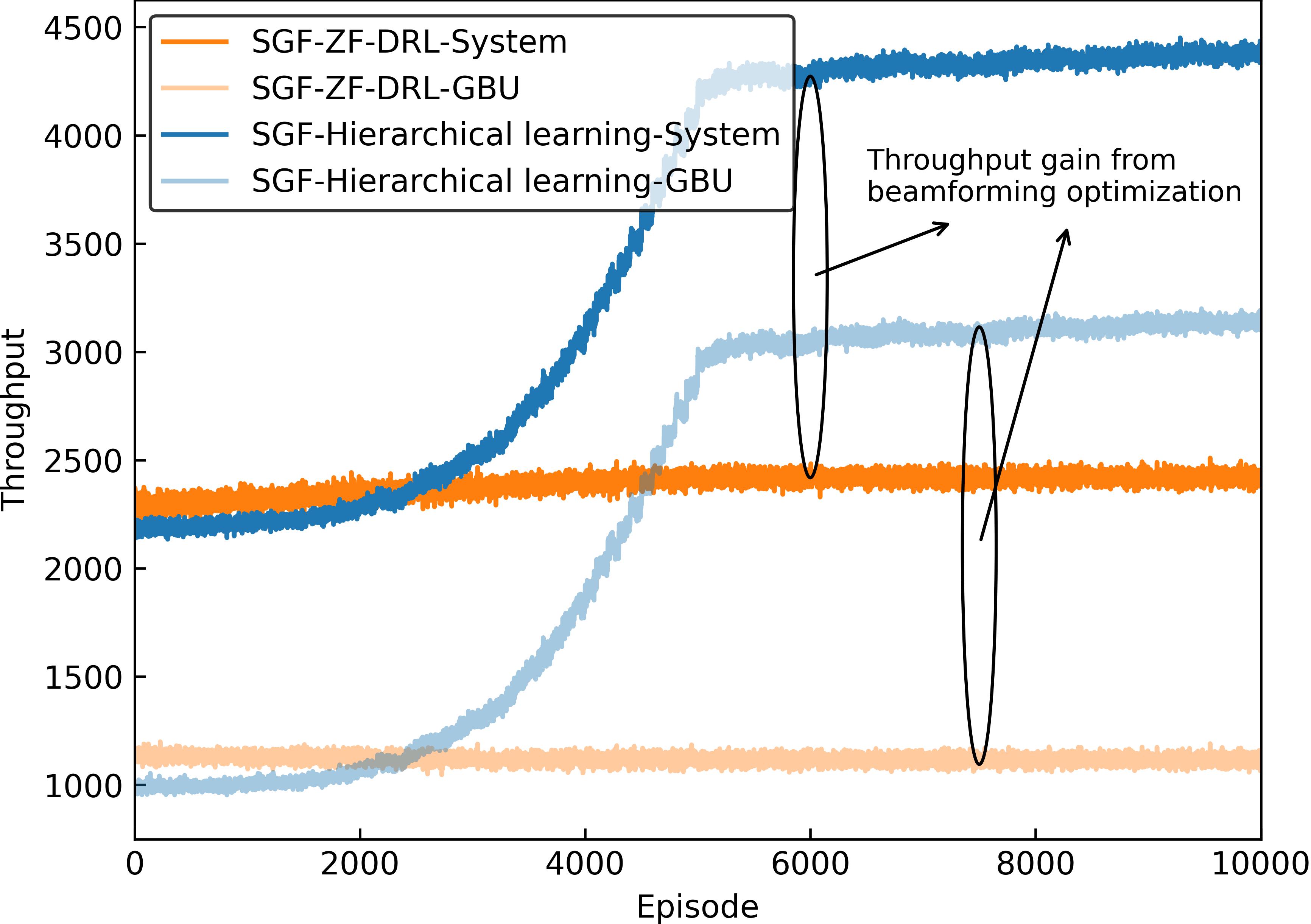}
        \caption{Throughput gain.}
        \label{Throughput gain.}
    \end{subfigure}
    \begin{subfigure}{0.45\linewidth}
        \includegraphics[width=\linewidth]{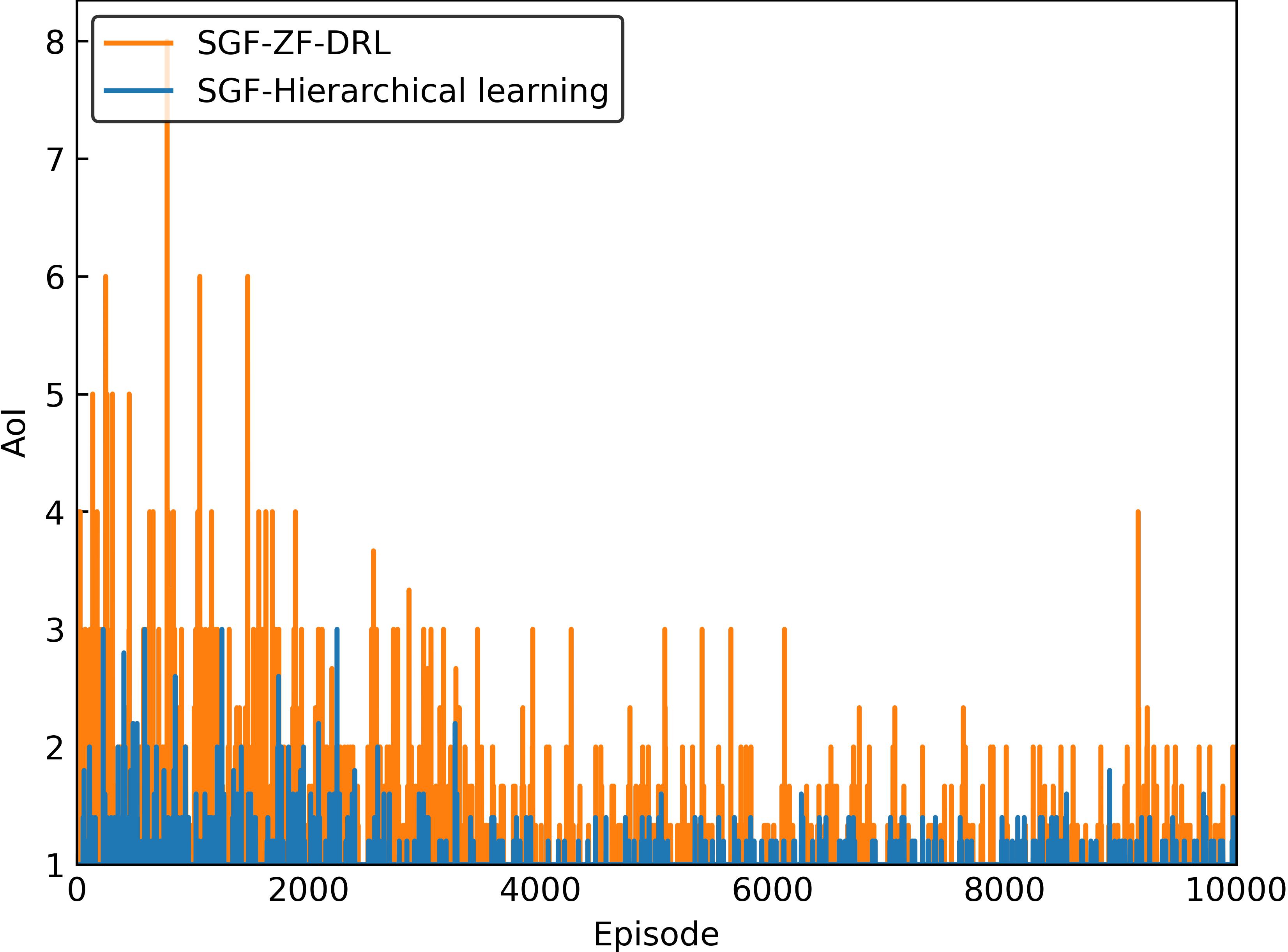}
        \caption{AoI gain.}
        \label{AoI gain}
    \end{subfigure}
    \caption{Performance gain from beamforming optimization.}
    \label{Performance Comparison.}
\end{figure*}

\subsubsection{Performance Comparison}
In Fig.~\ref{Performance Comparison.}, we study the joint optimization performance of HL by comparing it with an OMA scheme, where the beamforming is implemented by zero-forcing pre-coder (ZF) and the scheduling policy is adapted by RL. It can be observed that, under both NOMA and OMA, the SGF framework is able to support additional transmission tasks from GFUs. As shown in Fig.~\ref{Performance Comparison.}(\subref{Throughput gain.}), a $31.82 \%$ troughput gain is achieved over ZF by the optimized beamforming from the upper-level agent. We can also notice that the optimized beamforming is also facilitating the GF transmissions in the lower-level task, as shown in Fig.~\ref{Performance Comparison.}(\subref{AoI gain}). This is because an optimized beamforming enable efficient transmissions of GBUs, thus providing more transmission opportunities for GFUs, which result in better AoI performance.

\subsubsection{Performance with Growing Number of GFUs}
\begin{figure}[ht]
    \centering
    \begin{subfigure}{0.45\linewidth}
        \includegraphics[width=\linewidth]{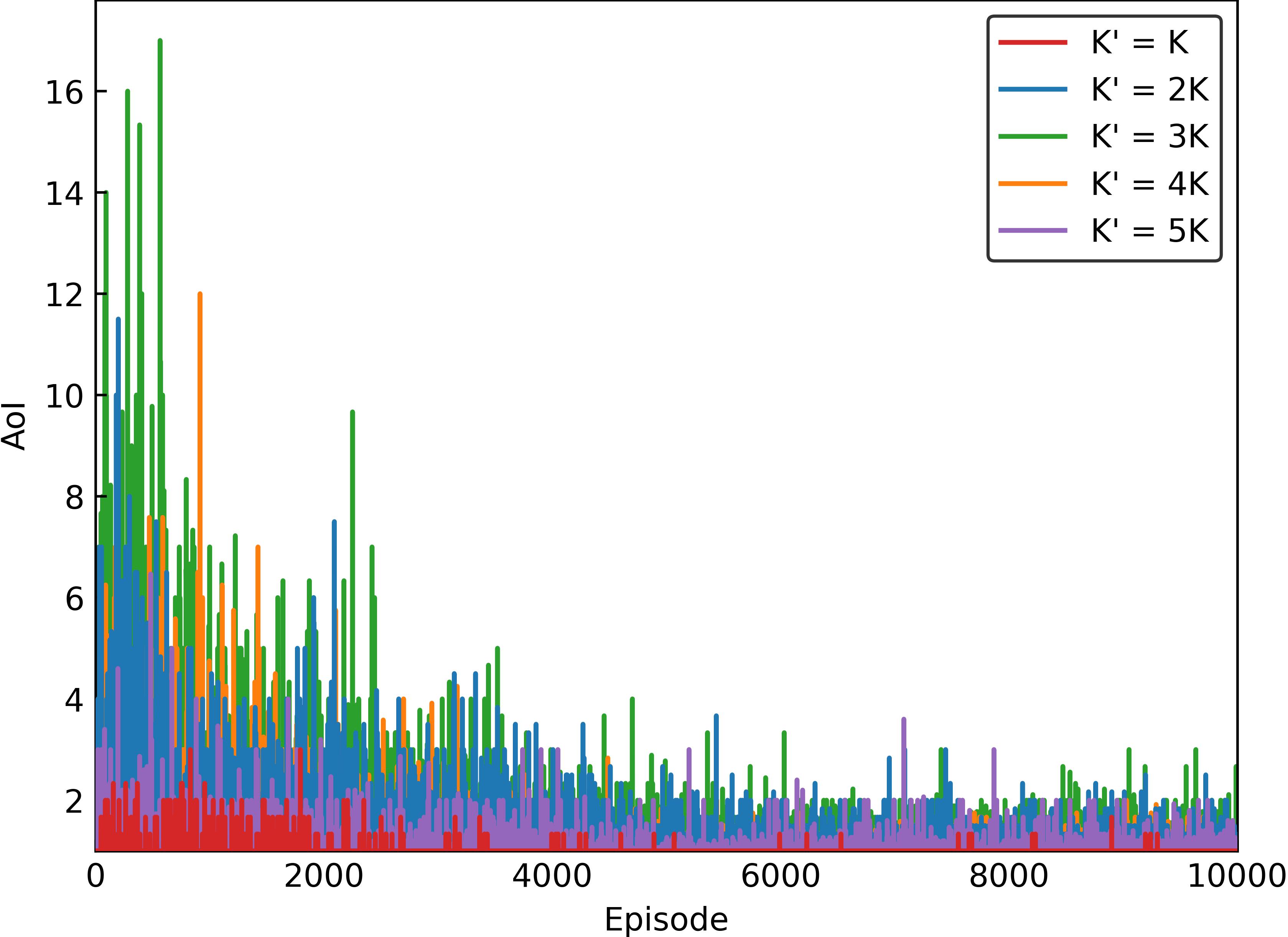}
        \label{K=3}
        \caption{K = 3}
    \end{subfigure}
    \begin{subfigure}{0.45\linewidth}
        \includegraphics[width=\linewidth]{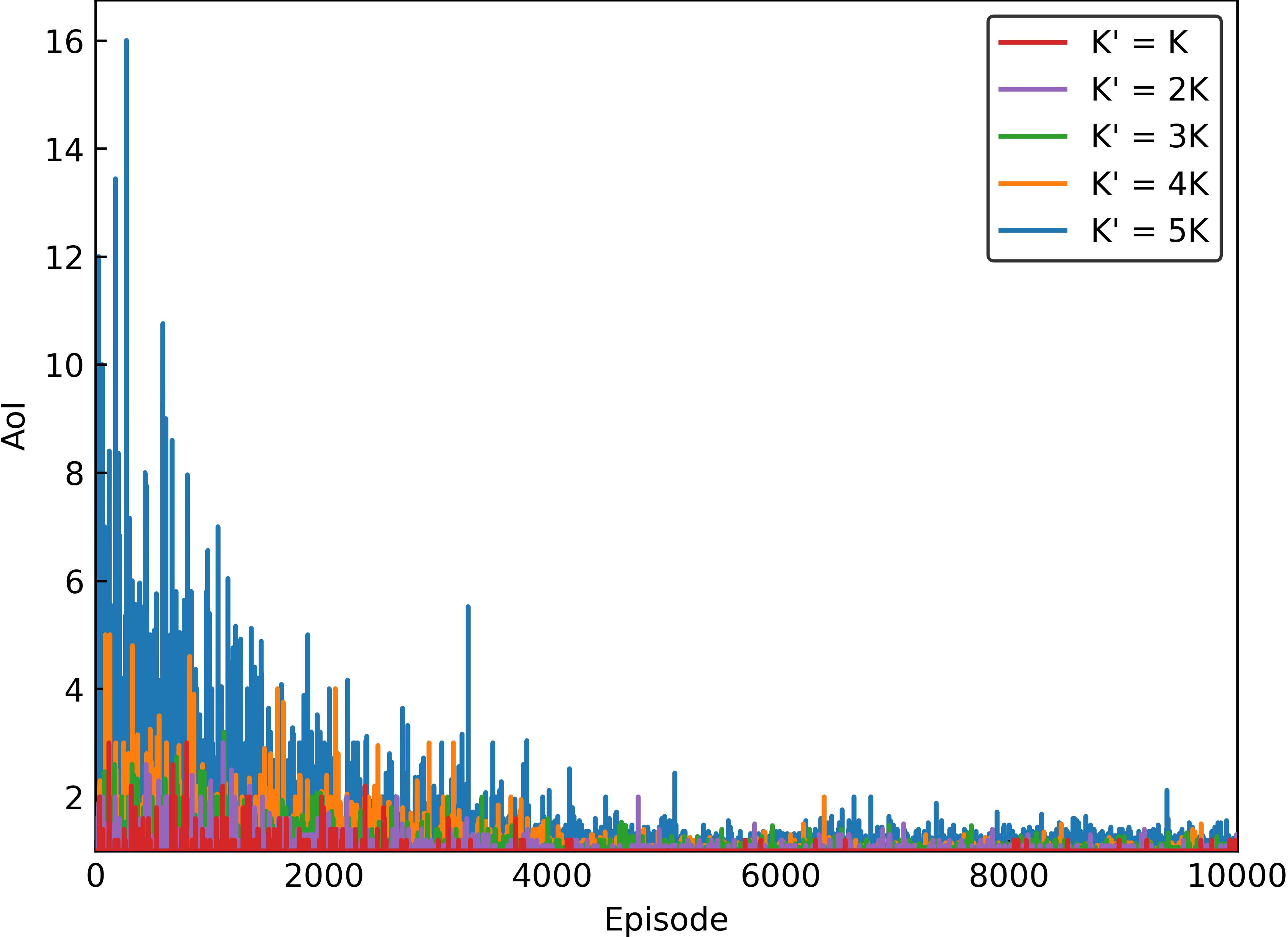}
        \label{K=5}
        \caption{K = 5}
    \end{subfigure}
    \caption{AoI performance over growing GFU counts.}
    \label{AoI performance over growing GFU counts.}
\end{figure}

In this subsection, we investigate the performance of the proposed algorithm over the growing GFU counts from $K' = K$ to $K' = 5K$, where $K=\{3, 5\}$, and $K'= \{K, 2K, 3K, 4K, 5K\}$. As illustrated in Table~\ref{Throughput gain with increasing GFU counts.}, compared to the baseline of $K' = K$, the NOMA-assisted SGF system is able to achive higher system capacity, where maxim throughput gains of $53.48 \%$ and $108.97\%$ are achieved when $K' = 5K$. In the meantime, the proposed hierarchical learning is able to find the optimal scheduling scheme, as shown in Fig.~\ref{AoI performance over growing GFU counts.}.

\begin{table*}[htbp]
\centering
\caption{Throughput gain with increasing GFU counts.}
\label{Throughput gain with increasing GFU counts.}
\begin{tabular}{|c|c|c|c|c|c|c|}
\hline
\multicolumn{2}{|c|}{$K'$}                           & $K$     & $2K$     & $3K$     & $4K$     & $5K$     \\ \hline
\multirow{2}{*}{$K = 3$} & Throughput             & 3158.30 & 3444.25 & 3681.02 & 4578.21 & 4847.24 \\ \cline{2-7}
                         & Throughput gain        & baseline & +9.05\%  & +16.55\% & +44.96\% & +53.48\% \\ \hline
\multirow{2}{*}{$K = 5$} & Throughput             & 2945.84 & 3417.81 & 3878.89 & 5003.11 & 6155.86 \\ \cline{2-7}
                         & Throughput improvement & baseline & +16.02\% & +31.67\% & +69.84\% & +108.97\% \\ \hline
\end{tabular}
\end{table*}



\section{Conclusion}\label{Conclusion}
The joint optimization of transmission scheduling and beamforming design was investigated to maintain data freshness and ensure high system capacity. More particularly, firstly, a NOMA-assisted SGF framework was formulated to support GFUs by utilizing residual resources from GBUs. Secondly, a hierarchical learning algorithm was proposed to decouple the formulated problem into two sequential subproblems, where the upper-level network is optimized to serve GBUs efficiently and maximize available transmission slots for the lower-level process. While the lower-level network is adapted to promote GFUs' transmission and avoid collisions. Simulation results demonstrated that the proposed NOMA-assisted SGF framework achieves higher capacity without bandwidth expansion. Moreover, the hierarchical learning algorithm achieves approximately a 31.82\% gain while maintaining information fresh.

\bibliographystyle{IEEEtran}
\bibliography{conference_101719}
\end{document}